\documentstyle[12pt,bezier]{article}
\begin{document}
\def \inbar{\vrule height1.5ex width.4pt depth0pt}
\def \xC{\relax\hbox{\kern.25em$\inbar\kern-.3em{\rm C}$}}
\def \xR{\relax{\rm I\kern-.18em R}}
\newcommand{\R}{\xR}
\newcommand{\C}{\xC}
\newcommand{\xZ}{Z \hspace{-.08in}Z}
\newcommand{\Z}{Z \hspace{-.08in}Z}
\newcommand{\xbe}{\begin{equation}}
\newcommand{\be}{\begin{equation}}
\newcommand{\xee}{\end{equation}}
\newcommand{\ee}{\end{equation}}
\newcommand{\xbea}{\begin{eqnarray}}
\newcommand{\bea}{\begin{eqnarray}}
\newcommand{\xeea}{\end{eqnarray}}
\newcommand{\eea}{\end{eqnarray}}
\newcommand{\xnn}{\nonumber}
\newcommand{\nn}{\nonumber}
\newcommand{\xkt}{\rangle}
\newcommand{\kt}{\rangle}
\newcommand{\xbr}{\langle}
\newcommand{\br}{\langle}
\newcommand{\xcun}{\mbox{\footnotesize${\cal N}$}}
\newcommand{\cun}{\mbox{\footnotesize${\cal N}$}}
\newcommand{\cum}{\mbox{\footnotesize${\cal M}$}}
\newcommand{\Q}{{\cal Q}}
.
\vspace{4cm}

\begin{center}
{\Huge {\bf Quantum Mechanical Symmetries and
Topological Invariants}}\\
\vspace{1cm}
{\large{\bf K.~Aghababaei~Samani and A.~Mostafazadeh}}
\end{center}
\vspace{5cm}
{\bf Corresponding Author:} Ali Mostafazadeh\\
{\bf Address:} Department of Mathematics, Ko\c{c} University, Rumeli Feneri Yolu,
 80900 Istanbul, TURKEY.\\
{\bf E-mail:} amostafazadeh@ku.edu.tr

\newpage

\title{Quantum Mechanical Symmetries and Topological~Invariants}
\author{K.~Aghababaei Samani$ ^{a}$ and
A.~Mostafazadeh$ ^{b}$\\ \\
$ ^{a}$~Institute for Advanced Studies in Basic Sciences,\\
45195-159 Gava Zang, Zanjan, IRAN\thanks{E-mail address: 
samani@iasbs.ac.ir}\\
$ ^{b}$~Department of Mathematics, Ko\c{c} University,\\
Rumelifeneri Yolu, 80900 Istanbul, TURKEY\thanks{E-mail address: 
amostafazadeh@ku.edu.tr}}
\date{ }
\maketitle

\begin{abstract}
We give the definition and explore the algebraic structure of a class of quantum
symmetries, called topological symmetries, which are generalizations of 
supersymmetry in the sense that they involve topological invariants similar to the 
Witten index. A topological symmetry (TS) is specified by an integer $n>1$, which
determines its grading properties, and an $n$-tuple of positive integers 
$(m_1,m_2,\cdots,m_n)$. We identify the algebras of supersymmetry, 
$p=2$ parasupersymmetry, and fractional supersymmetry of order $n$ with 
those of the $\Z_2$-graded TS of type $(1,1)$, $\Z_2$-graded TS of type 
$(2,1)$, and $\Z_n$-graded TS of type $(1,1,\cdots,1)$, respectively. We also
comment on the mathematical interpretation of the topological invariants 
associated with the $\Z_n$-graded TS of type $(1,1,\cdots,1)$. For $n=2$, the
invariant is the Witten index which can be identified with the analytic index of 
a Fredholm operator. For $n>2$, there are $n$ independent integer-valued
invariants. These can be related to differences of the dimension of the kernels
of various products of $n$ operators satisfying certain conditions. 
\end{abstract}


\baselineskip=24pt

\section{Introduction} 
Supersymmetric quantum mechanics was originally introduced as a toy 
model used to study some of the features of supersymmetric field theories 
\cite{witten-82}. This simple toy model has, however, proven to be a very 
useful tool in dealing with a variety of problems in quantum and statistical 
mechanics \cite{susy,review,junker}. Supersymmetric quantum mechanics has
also been used to derive some of the very basic results of differential topology.
Among these are the supersymmetric derivation of the Morse 
inequalities \cite{morse} and supersymmetric proofs of the Atiyah-Singer index
theorem \cite{susy-index}.

The relationship between supersymmetry and topological invariants such as
the indices of the elliptic operators is our main motivation for seeking general 
quantum mechanical symmetries with topological properties similar to those of 
supersymmetry.

Various generalizations of supersymmetry have been considered in the literature
\cite{ru-sp,para,def,frac,frac2,ex-ge}. Among these only extended and generalized
supersymmetries \cite{ex-ge} and a certain class of $p=2$ parasupersymmetries 
\cite{ijmpa-97} are known to share the topological characteristics of supersymmetry. 

The strategy pursued in this article is as follows. First, we introduce the notion of
a topological symmetry (TS) by formulating a set of basic principles that ensure the desired
topological properties. Then, we investigate the underlying algebraic structure of 
these symmetries. This is necessary for seeking a mathematical interpretation of the 
corresponding topological invariants. 

We have recently reported our preliminary results on $\Z_2$-graded TSs of type
$(1,1)$ and $(2,1)$ in \cite{p34a}. The purpose of the present article is to 
generalize the results of \cite{p34a} to arbitrary $\Z_n$-graded TSs.
 
The organization of the article is as follows. In section~2, we give the definition
of a general $\Z_n$-graded TS and introduce the associated topological invariants.
In section~3, we consider the case of $n=2$ and derive the algebra of a 
$\Z_2$-graded TS of arbitrary type $(m_+,m_-)$. In particular, we show that for 
$m_-=1$, the algebra can be reduced to the algebra of supersymmetry or $p=2$ 
parasupersymmetry. In section~4, we consider the $\Z_n$-graded TSs. Here we 
discuss the properties of the grading operator and derive the algebra of  $\Z_n$-graded 
TSs of arbitrary type $(m_1,m_2,\cdots,m_n)$. In section~5, we comment on the 
mathematical interpretation of the topological invariants associated with TSs. In 
section~6, we study some concrete examples of quantum systems possessing TSs. 
In section~7, we summarize our results and present our concluding remarks. The 
appendix includes the proof of some of the mathematical results that we use in our 
analysis.

\section{$\Z_n$-Graded TSs}

In order to describe the concept of a topological symmetry, we first give some 
basic definitions. In the following we shall only consider the quantum systems 
with a self-adjoint Hamiltonian $H$. We shall further assume that all the energy
levels are at most finitely degenerate.

\begin{itemize}
\item[~] {\bf Definition~1:} Let $n$ be an integer greater than 1, ${\cal H}$
denote the Hilbert space of a quantum system, and ${\cal H}_1,{\cal H}_2,\cdots,
{\cal H}_n$ be (nontrivial) subspaces of ${\cal H}$. Then a state vector is said 
to have {\em definite color} $c_\ell$ iff it belongs to ${\cal H}_\ell$.
\item[~] {\bf Definition~2:} A quantum system is said to be {\em $\Z_n$-graded}
iff its Hilbert space is the direct sum of $n$ of its (nontrivial) subspaces 
${\cal H}_\ell$, and its Hamiltonian has a complete set of eigenvectors with definite
color.
\item[~] {\bf Definition~3:} Let $m_\ell$ be positive integers for all 
$\ell\in\{1,2,\cdots,n\}$, and $m:=\sum_{\ell=1}^nm_\ell$. Then a quantum system
is said to possess a $\Z_n$-{\em graded topological symmetry of type} 
$(m_1,m_2,\cdots, m_n)$ iff the following conditions are satisfied.
	\begin{itemize}
	\item[a)] The quantum system is $\Z_n$-graded;
	\item[b)] The energy spectrum is nonnegative;
	\item[c)] For every positive energy eigenvalue $E$, there is a positive
        integer $\lambda_E$ such that $E$ is $\lambda_E m$-fold degenerate, and
	the corresponding eigenspaces are spanned by $\lambda_Em_1$ vectors of 
	color $c_1$, $\lambda_Em_2$ vectors of color $c_2$, $\cdots$, and 
	$\lambda_Em_n$ vectors of color $c_n$.
	\end{itemize}
\item[~] {\bf Definition~4:} A topological symmetry is said to be {\em uniform} iff
        for all positive energy eigenvalues $E$, $\lambda_E=1$. 
\item[~] {\bf Theorem~1:} Consider a quantum system possessing a $\Z_n$-graded 
topological symmetry of type $(m_1, m_2, \cdots, m_n)$, and let  $n^{(0)}_\ell$ 
denote the number of zero-energy states of color $c_\ell$. Then for all 
$i,j\in\{1,2,\cdots,n\}$, the integers 
	\be
	\Delta_{ij}:=m_in^{(0)}_j-m_jn^{(0)}_i
	\label{1}
	\ee
remain invariant under continuous symmetry-preserving changes of the 
quantum system.\footnote{Here we consider quantum systems whose energy 
spectrum depends on a set of continuous parameters. These parameters may
be identified with coupling constants or geometric quantities entering 
the definition of the Hamiltonian and/or the Hilbert space. A continuous
change of the system corresponds to a continuous change of these parameters.}
\item[~] {\bf Proof:} The proof of this theorem is essentially the same as the 
proof of the topological invariance of the Witten index of supersymmetry. A 
symmetry-preserving continuous change of the quantum system will preserve 
the particular degeneracy and grading structures of positive energy levels. 
Because under such a change an initial zero energy eigenstate can only become
a positive energy eigenstate and vice versa, the only possible change in the 
number of zero energy eigenstates are the ones involving the changes of 
$n^{(0)}_i$ of the form
	\be
	n^{(0)}_i \to \tilde{n}^{(0)}_i:= n^{(0)}_i+k~m_i\;,
	\label{2}
	\ee
where $k$ is an integer greater than or equal to $-n^{(0)}_i/m_i$. Moreover, 
such a change must occur simultaneously for all $n^{(0)}_i$'s, i.e., the 
transformation~(\ref{2}) is valid for all $i\in\{1,2,\cdots,n\}$. Therefore under 
such a symmetry-preserving change of the system,
	\[\Delta_{ij}\to\tilde  \Delta_{ij}:=m_i\tilde n^{(0)}_j-
	m_j\tilde n^{(0)}_i=m_i(n^{(0)}_j+km_j)-m_j(n^{(0)}_i+km_i)
	=m_in^{(0)}_j-m_jn^{(0)}_i=\Delta_{ij}\;,\]
i.e., $\Delta_{ij}$'s remain invariant. $\Box$
\end{itemize}

A direct consequence of Theorem~1 is that (the value of) any function of 
$\Delta_{ij}$'s is a topological invariant of the system. In particular,  
$\Delta_{ij}$'s  are the basic topological invariants. A typical example of a 
derived invariant is 
	\be
	\Delta:=\frac{1}{2}\,\sum_{i,j=1}^n(\Delta_{ij})^2\;.
	\label{D}
	\ee
Note that $\Delta$ is a measure of the existence of the zero-energy states.
 
Let us next observe that the topological symmetries are simple 
generalizations of supersymmetry. First we recall that the Hilbert space of a 
supersymmetric system is $\Z_2$-graded. We can relate the $\Z_2$-grading of
the Hilbert space ${\cal H}$ to the existence of  a `parity' or `grading' operator 
$\tau:{\cal H}\to{\cal H}$ satisfying
	\bea
	\tau^2&=&1\;,
	\label{t1}\\
	\tau^\dagger&=&\tau\;,
	\label{t2}\\
	\left[H,\tau\right]&=&0\;.
	\label{t3}
	\eea
Here we use a $\dagger$ to denote the adjoint of the corresponding operator. We 
can identify the subspaces ${\cal H}_1$ and ${\cal H}_2$ with the eigenspaces
of $\tau$,
	\be
	{\cal H}_1={\cal H}_+\;,~~~
	{\cal H}_2={\cal H}_-\;,~~~
	{\cal H}_\pm:=\{\psi\in{\cal H}~|~\tau\psi=\pm\psi\}\;.
	 \label{3}
	\ee
Now, consider the superalgebra
	\bea
	[H,\Q]&=&0\;,
	\label{s1}\\
	\frac{1}{2}\{\Q,\Q^\dagger\}&=&H\;,
	\label{s2}\\
	\Q^2&=&0\;,
	\label{s3}
	\eea
of supersymmetric quantum mechanics with one nonself-adjoint symmetry 
generator $\Q$ satisfying 
	\be
	\{\Q,\tau\}=0\;.
	\label{q1}
	\ee
It is well-known that using the superalgebra (\ref{s1}) -- (\ref{s3}) together 
with the properties of the grading operator (\ref{t1}) -- (\ref{t3}) and the 
symmetry generator (\ref{q1}), one can show that the conditions a) -- c) of 
Definition~3, with $n=2$ and $m_1=m_2=1$, are satisfied. Therefore, 
supersymmetry is a $\Z_2$-graded TS of type $(1,1)$. For a $\Z_2$-graded 
TS of type $(1,1)$, there is a single basic topological invariant namely 
$\Delta_{11}$. This is precisely the Witten index.

\section{Algebraic Structure of  $\Z_2$-Graded TSs}

In this section we shall explore the algebraic structure of the $\Z_2$-graded 
TSs that fulfil the following conditions.
	\begin{itemize}
	\item[--] The $\Z_2$-grading is achieved by a grading operator $\tau$
	satisfying (\ref{t1}) -- (\ref{t3});
	\item[--] There is a single nonself-adjoint symmetry generator $\Q$;
	\item[--] $\Q$ is an odd operator, i.e., it satisfies Eq.~(\ref{q1}).
	\end{itemize}
We shall only treat the case of the uniform TSs. The algebraic structure of 
nonuniform TSs is easily obtained from that of the uniform topological symmetries
(UTSs). In fact, the algebraic relations defining uniform and nonuniform TSs
of the same type are identical.
 
In order to obtain the algebraic structures that support TSs, we shall use the 
information on the degeneracy structure of the corresponding systems and the 
properties of the grading operator and the symmetry generator to construct matrix 
representations of the relevant operators in the energy eigenspaces ${\cal H}_E$ 
with positive eigenvalue $E$. We shall use the notation $O^E$ for the restriction 
of an operator $O$ onto the eigenspace ${\cal H}_E$. Throughout this article $E$ 
stands for a positive energy eigenvalue. The zero-energy eigenspace (kernel of $H$)
is denoted by ${\cal H}_0$.

In view of Eqs.~(\ref{t3}) and (\ref{s1}), $\tau^E$ and $\Q^E$ are 
$m\times m$ matrices acting in ${\cal H}_E$. We also have the trivial
identity: $H^E=E~I_m$, where $I_m$ denotes the $m\times m$ unit 
matrix. 

Next, we introduce the self-adjoint symmetry generators
	\be
	Q_1:=\frac{1}{\sqrt{2}}\:(\Q+\Q^\dagger)\;~~~
	{\rm and}~~~Q_2:=\frac{-i}{\sqrt{2}}\:(\Q-\Q^\dagger)\;
	\label{4}
	\ee
where $i:=\sqrt{-1}$. Note that because $\tau$ is self-adjoint, we have
	\be
	\{Q_j,\tau\}=0~~~{\rm for}~~~j\in\{1,2\}.
	\label{qt}
	\ee

Now in view of Eq.~(\ref{t3}), we can choose a basis in ${\cal H}_E$ in 
which $\tau$ is diagonal. Then using Eqs.~(\ref{t1}), (\ref{4}), (\ref{qt}),
and the self-adjointness of $Q_j$, we obtain the following matrix 
representations for $\tau^E$, $Q_j^E$, and $\Q^E$.
	\bea
	\tau^E&=&{\rm diag}(\underbrace{1,1,\cdots,1}_{m_+ {\rm times}},
	\underbrace{-1,-1,\cdots,-1}_{m_- {\rm times}})\;,
	\label{t-rep}\\
	Q_j^E&=&\left(\begin{array}{cc}
	0&A_j\\
	A^\dagger_j&0
	\end{array}\right)\;,
	\label{q-rep}\\
	\Q^E&=&\frac{1}{\sqrt{2}}\,\left(\begin{array}{cc}
	0&A_1+iA_2\\
	A^\dagger_1+iA^\dagger_2&0
	\end{array}\right)\;,
	\label{Q-rep}
	\eea
where `diag$(\cdots)$' stands for a diagonal matrix with diagonal entries `$\cdots$',
$0$'s denote appropriate zero matrices, and $A_j$ are $m_+\times m_-$ complex 
matrices.

The next step is to find general identities satisfied by $Q_j^E$ and $\Q^E$
for all $E>0$. In order to derive the simplest such identities we appeal to
the Cayley-Hamilton theorem of  linear algebra. This theorem states that an 
$m\times m$ matrix $Q$ satisfies its
characteristic equations, ${\cal P}_Q(Q)=0$, where ${\cal P}_Q(x)
:=\det(xI_m-Q)$ is the characteristic polynomial for $Q$. Using this theorem
we can prove the following lemma. The proof is given in the appendix.
	\begin{itemize}
	\item[~] {\bf Lemma~1:} Let $m_\pm$ be positive integers, $m:=m_++
	m_-$, and $Q$ be an $m\times m$ matrix of the form:
		\be
		Q=\left(\begin{array}{cc}
		0&X\\
		Y&0
		\end{array}\right)\;,
		\label{xy}
		\ee
	where $X$ and $Y$ are $m_+\times m_-$ and $m_-\times m_+$ 
	complex matrices. Let ${\cal P}_{XY}(x)$ and ${\cal P}_{YX}(x)$
	denote the characteristic polynomials for $XY$ and $YX$, respectively.
	Then ${\cal P}_{YX}(Q^2)Q={\cal P}_{XY}(Q^2)Q=0$. Furthermore,
	if $m_+=m_-$, then ${\cal P}_{YX}(Q^2)={\cal P}_{XY}(Q^2)=0$.
        \end{itemize}

Applying this lemma to $Q_j^E$ and $\Q^E$, we find for $m_+=m_-$
	\bea
	{\cal P}_j[(Q_j^E)^2]&=&0\;,
	\label{9}\\
	{\cal P}[(\Q^E)^2]&=&0\;,
	\label{10}
	\eea
and for $m_+>m_-$
	\bea
	{\cal P}_j[(Q_j^E)^2]Q^E_j&=&0\;,
	\label{11}\\
	{\cal P}[(\Q^E)^2]\Q^E&=&0\;,
	\label{12}
	\eea
where ${\cal P}_j(x)$ and ${\cal P}(x)$ denote the characteristic polynomials 
of $A_j^\dagger A_j$ and $(A_1^\dagger+iA_2^\dagger)(A_1+iA_2)/2$, 
respectively,
	
For $m_->m_+$, the roles of $A_j$ and $A_j^\dagger$ are interchanged.
We shall, therefore, restrict our attention to the case where $m_+\geq m_-$.

We can write Eqs.~(\ref{9}) -- (\ref{12}) in terms of the roots of the 
characteristic polynomials ${\cal P}_j(x)$ and ${\cal P}(x)$. This yields
	\bea
	\left[(Q^E_j)^2 - \mu_{j1}^EI_m\right]
	\left[(Q^E_j)^2 - \mu_{j2}^EI_m\right]
	\cdots \left[(Q^E_j)^2 - \mu_{jm_-}^EI_m\right]
	(Q^E_j)^{1-\delta(m_+,m_-)}&=&0\;,
	\label{13}\\
	\left[(\Q^E)^2 - \kappa_{1}^EI_m\right]
	\left[(\Q^E)^2 - \kappa_{2}^EI_m\right]
	\cdots \left[(\Q^E)^2 - \kappa_{m_-}^EI_m\right]
	(\Q^E)^{1-\delta(m_+,m_-)}&=&0\;,
	\label{14}
	\eea
where $\mu_{j\ell}^E$ and $\kappa_\ell^E$ are the roots\footnote{Note that 
the roots are not necessarily distinct.}  of ${\cal P}_j(x)$ and ${\cal P}(x)$, 
respectively, and
	\[\delta(m_+,m_-)=\delta_{m_+,m_-}:=\left\{\begin{array}{ccc}
	1&{\rm for}& m_+=m_-\\
	0&{\rm for}& m_+\neq m_-\;.\end{array}\right.\]

In order to promote Eqs.~(\ref{13}) and (\ref{14}) to operator relations, we 
introduce the operators $M_{j\ell}$ and ${\cal K}_\ell$ (for each $j\in\{1,2\}$
and $\ell\in\{1,2,\cdots,m_-\}$)  which commute with $H$ and have the 
representations:
	\be
	M_{j\ell}^E=\mu_{j\ell}^E I_m~~~~{\rm and}~~~~
	{\cal K}_\ell^E=\kappa_\ell^E I_m
	\label{17}
	\ee
in ${\cal H}_E$. We then deduce from Eqs.~(\ref{13}), (\ref{14}) and 
(\ref{17}) that $M_{j\ell}$ and ${\cal K}_\ell$ must commute with $\tau$ and
$Q_j$. Furthermore, they should satisfy the algebra
	\bea
	(Q_j^2 - M_{j1})(Q_j^2 - M_{j2})\cdots(Q_j^2 - M_{jm_-})
	Q_j^{1-\delta(m_+,m_-)}&=&0\;,
	\label{a1}\\
	(\Q^2 - {\cal K}_1)(\Q^2 - {\cal K}_2)\cdots(\Q^2 - {\cal K}_{m_-})
	\Q^{1-\delta(m_+,m_-)}&=&0\;.
	\label{a2}
	\eea

Note that the roots $\kappa^E_\ell$ and $\mu^E_{j\ell}$ are defined using the 
matrices $A_j$. This suggests that the operators $M_{j\ell}$ and 
${\cal K}_\ell$ are not generally independent. Furthermore, because 
$A^\dagger_jA_j$ are Hermitian matrices, the roots $\mu^E_{j\ell}$, which 
are in fact the eigenvalues of $A^\dagger_jA_j$, are real. This in turn suggests
that the operators $M_{j\ell}$ are self-adjoint.

In summary, the algebra of general $\Z_2$-graded topological symmetry of type
$(m_+,m_-)$ which is generated by one odd nonself-adjoint generator $\Q$ is 
given by Eqs.~(\ref{a1}) and (\ref{a2}) where $m_+$ is assumed (without loss
of generality) not to be smaller than $m_-$, the operators $M_{j\ell}$ and 
${\cal K}_\ell$ commute with $H$, $\tau$ and $\Q$, and $M_{j\ell}$ are 
self-adjoint. Moreover, $M_{j\ell}$ and ${\cal K}_\ell$  have similar degeneracy 
structure as the Hamiltonian\footnote{The degeneracy structure of  these operators
will be the same as that of the Hamiltonian, if their eigenvalues $\mu_{j\ell}^E$ 
and $\kappa_\ell^E$ are distinct for different $E$.} (at least for positive energy 
eigenvalues). In particular, it might be possible to express $H$ as a function of 
$M_{j\ell}$ and ${\cal K}_\ell$.

In order to elucidate the role of the operators $M_{j\ell}$ and ${\cal K}_\ell$ 
and their relation to the Hamiltonian, we shall next consider the $\Z_2$-graded 
UTSs of type $(m_+,1)$.

If $m_-=1$, then $A_j^\dagger A_j$ and 
$(A_1^\dagger+iA_2^\dagger)(A_1+iA_2)/2$ are respectively real and 
complex scalars. In this case, ${\cal P}_j(x)=x-\mu^E_j$ and ${\cal P}(x)=
x-\kappa^E$, where 
	\be
	\mu^E_j=A_j^\dagger A_j\;,~~~~~
	\kappa^E=\frac{1}{2}\, (A_1^\dagger+iA_2^\dagger)(A_1+iA_2)=
	\frac{1}{2}\,[(A_1^\dagger A_1-A_2^\dagger A_2)+
	i(A_1^\dagger A_2+A_2^\dagger A_1)]\;,
	\label{20}
	\ee
and the algebra~(\ref{a1}) and (\ref{a2}) takes the form
	\bea
	(Q_j^2 - M_{j}) Q_j^{1-\delta(m_+,1)}&=&0\;,
	\label{a1-1}\\
	(\Q^2 - {\cal K})\Q^{1-\delta(m_+,1)}&=&0\;.	
	\label{a1-2}
	\eea
Here we have used the abbreviated notation: $M_j=M_{j1}$ and 
${\cal K}={\cal K}_1$. 

Next, we define the self-adjoint operators 
	\be
	K_1={\cal K}+{\cal K}^\dagger~~~{\rm and}~~~
	K_2=-i({\cal K}-{\cal K}^\dagger)\;.
	\label{kk}
	\ee
In view of Eqs.~(\ref{17}) and (\ref{20}) we have
	\be
	M_2=M_1-K_1\;.
	\label{21}
	\ee
In the following we shall consider the cases $m_+=1$ and 
$m_+>1$ separately.

\subsection{$\Z_2$-Graded  UTS of Type $(1,1)$}

Setting $m_+=1$ in Eqs.~(\ref{a1-1}) and (\ref{a1-2}), we find
	\bea
	Q_j^2&=&M_j\;,
	\label{g1}\\
	\Q^2&=&{\cal K}\;.
	\label{g2}
	\eea
If we express $\Q$ in terms of $Q_j$ and use Eqs.~(\ref{kk}) and (\ref{21}), 
we can write Eqs.~(\ref{g1}) and (\ref{g2}) in the form
	\bea
	Q_1^2&=&M_1\;,
	\label{h1}\\
	Q_2^2&=&M_1-K_1\;,
	\label{h2}\\
	\{Q_1,Q_2\}&=&K_2\;.
	\label{h3}
	\eea

Now, we observe that Eqs.~(\ref{h1}) -- (\ref{h3}) remain form-invariant 
under the linear transformations of the form
	\be
	\begin{array}{c}
	Q_1\to\tilde Q_1=a~Q_1+b~Q_2\;,\\
	Q_2\to\tilde Q_2=c~Q_1+d~Q_2\;,
	\end{array}
	\label{trans}
	\ee
where $a,~b,~c$ and $d$ are self-adjoint operators commuting with all other 
operators. More specifically, $\tilde Q_j$ satisfy Eqs.~(\ref{h1}) -- (\ref{h3}) 
provided that $M_1$ and $K_j$ are transformed according to
	\bea
	M_1&\to&\tilde M_1:=(a^2+b^2)M_1-b^2K_1+abK_2\;,
	\label{tt1}\\
	K_1&\to&\tilde K_1:=(a^2+b^2-c^2-d^2)M_1++(d^2-b^2)K_1+
	(ab-cd)K_2\;,
	\label{tt2}\\
	K_2&\to&\tilde K_2:=2(ac+bd)M_1-2bd K_1+(ad+bc)K_2\;.
	\label{tt3}
	\eea
In particular, there are transformations of the form~(\ref{trans}) for which 
$\tilde K_j=0$. These correspond to the choices for $a,~b,~c$ and $d$ that 
satisfy (either of)
	\be
	\frac{a+ic}{b+id}=-\frac{K_2}{2M_1}\pm i
	\sqrt{1-\frac{K_1}{M_1}-
	\frac{K_2^2}{4M_1^2}}\;.
	\label{tt4}
	\ee
One can use the representations of $K_j$ and $M_1$ in the eigenspaces
${\cal H}_E$ to show that the terms in the square root in (\ref{tt4}) yield a 
positive self-adjoint operator, provided that the kernel of $M_1$ is a subspace
of the zero-energy eigenspace ${\cal H}_0$. 

The above analysis shows that we can reduce the general algebra
(\ref{h1}) -- (\ref{h3}) to the special case where $K_j=0$. Writing this 
algebra in terms of $\Q$, we obtain the superalgebra (\ref{s1}) -- (\ref{s3}) 
with $M_1$  replacing $H$. In other words, if we identify the Hamiltonian with
$M_1$, which we can always do, the algebra of $\Z_2$-graded topological 
symmetry of type $(1,1)$ reduces to that of supersymmetry.

\subsection{$\Z_2$-Graded UTSs of Type $(m_+,1)$ with $m+>1$}

If $M_+>1$, then Eqs.~(\ref{a1-1}) and (\ref{a1-2}) take the form
	\bea
	Q_j^3&=&M_j Q_j\;,
	\label{p1}\\
	\Q^3&=&{\cal K}\Q\;.	
	\label{p2}
	\eea
Again we express $\Q$ in terms of $Q_j$ and use Eqs.~(\ref{kk}) and 
(\ref{21}) to write (\ref{p1}) and (\ref{p2}) in the form
	\bea
	&&Q_1^3=M_1Q_1\;,
	\label{qq1}\\
	&&Q_2^3=(M_1-K_1)Q_2\;,
	\label{qq2}\\
	&&Q_2Q_1Q_2+\{Q_1,Q_2^2\}=(M_1-K_1)Q_1+K_2Q_2\;,
	\label{qq3}\\
	&&Q_1Q_2Q_1+\{Q_2,Q_1^2\}=M_1Q_2+K_2Q_1\;.
	\label{qq4}
	\eea
It is remarkable that these relations are also invariant under the transformations
(\ref{trans}) and (\ref{tt1}) -- (\ref{tt3}). Therefore, again we can reduce our
analysis to the special case where $K_j=0$. Substituting zero for $K_j$ in 
Eqs.~(\ref{qq1}) -- (\ref{qq4}), and writing them in terms of $\Q$, we obtain
	\bea
	&&[M_1,\Q]=0\;,
	\label{ps1}\\
	&&\{\Q^2Q^\dagger\}+\Q\Q^\dagger\Q=2M_1\Q\;,
	\label{ps2}\\
	&&\Q^3=0\;.
	\label{ps3}
	\eea
This is precisely the algebra of $p=2$ parasupersymmetry of Rubakov and 
Spiridonov \cite{ru-sp} with $H$ replaced by $M_1/2$.  Hence, if we identify 
$H$ with $M_1/2$, which we can always do, the algebra of $\Z_2$-graded 
topological symmetry of type $(m_+,1)$ with $m_+>1$ reduces to that of the
$p=2$ parasupersymmetry. 

As shown in Ref.~\cite{ijmpa-96a}, one can use the algebra (\ref{ps1}) --
(\ref{ps3}) of $p=2$ parasupersymmetry and properties of the grading operator
(\ref{t1}) -- (\ref{t3}) and (\ref{qt}) to obtain the general degeneracy structure
of a $p=2$ parasupersymmetric system. In general the algebra of $p=2$ 
parasupersymmetry does not imply the particular degeneracy structure of the 
$\Z_2$-graded UTS of type $(m_+,1)$, even for $m_+=2$. Therefore,
the $\Z_2$-graded UTS of type $(2,1)$ is a subclass of the general $p=2$ 
parasupersymmetries. As argued in Refs.~\cite{ijmpa-96a} and \cite{ijmpa-97},
these are parasupersymmetries for which an analog of the Witten index can be
defined. 

In Ref.~\cite{ijmpa-96a} it is also shown that the positive energy eigenvalues 
of a $p=2$ parasupersymmetric system can at most be triply degenerate, 
provided that the eigenvalues of $Q_1^E$ for all $E>0$ are nondegenerate. 
This means that the $\Z_2$-graded TSs of type $(m_+,1)$ with $m_+>2$ occur 
only if $Q_1^E$ have degenerate eigenvalues for all $E>0$. This suggests the 
presence of further (even) symmetry generators $L_a$ which would commute 
with $Q_1$ and label the basis eigenvectors within the degeneracy subspaces of
$Q_1$. The existence of these generators is an indication that the $\Z_2$-graded
TSs of type $(m_+,1)$ with $m_+>2$ are not uniform. 

\subsection{Special $\Z_2$-Graded TSs}

The analysis of the $\Z_2$-graded TSs of type $(m_+,1)$ shows that the 
corresponding algebras can be reduced to a simplified special case by a 
redefinition of the symmetry generators $Q_j$. This raises the question whether
this is also possible for the general $\Z_2$-graded TSs of type $(m_+,m_-)$. 
The reduction made in the case of $\Z_2$-graded TSs of type $(m_+,1)$ has
its roots in the form of the matrix representation of the corresponding symmetry
generators in ${\cal H}_E$. For a $\Z_2$-graded UTS of arbitrary type 
$(m_+,m_-)$, a similar reduction, which eliminates the operators 
${\cal K}_{\ell}$ in Eq.~(\ref{a2}), is possible, if we can find a transformation 
$Q_j\to\tilde Q_j$ which satisfies the following conditions.
	\begin{itemize}
	\item[1)] The transformed generators $\tilde Q_j$ have the representation
	\[\tilde Q_j^E=\left(\begin{array}{cc}
	0&\tilde A_j\\
	\tilde A_j^\dagger&0\end{array}\right)\]
	in ${\cal H}_E$.
	\item[2)] For all energy eigenvalues $E>0$, the corresponding matrices 
	$\tilde A_j$ which define $\tilde Q_j^E$ satisfy 
	$\tilde A_2=U \tilde A_1$, where $U$ is an $m_+\times m_+$ unitary 
	and anti-Hermitian matrix.
	\end{itemize}

The first condition is necessary for the invariance of the algebra (\ref{a1}) --
(\ref{a2}). It is satisfied by the linear transformations: $Q^E_j\to\tilde{Q}^E_j=
T_jQ_j^E T_j^\dagger$ where $T_j$ are $m\times m$ matrices of the form
	\[ T_j=\left(\begin{array}{cc}
	T_{j+}&0\\
	0&T_{j-}\end{array}\right)\;,\]
and $T_{j\pm}$ are $m_\pm\times m_\pm$ matrices. Under such a 
transformation $A_j$ transform according to $A_j\to\tilde A_j:=
T_{j+}A_jT_{j-}^\dagger$.

The second condition implies that the transformed matrices $\tilde A_j$ satisfy
	\be
	(\tilde A_1^\dagger+i\tilde A_2^\dagger)(\tilde A_1+i\tilde A_2)=
	A_1^\dagger (I_{m_+}+iU^\dagger)(I_{m_+}+iU)A_1
	= A_1^\dagger[(I_{m_+}-U^\dagger U)+i(U+U^\dagger)]A_1=0\;,
	\label{condi}
	\ee
where we have used the unitarity and anti-Hermiticity of $U$. The latter relation
indicates that for the transformed system $\kappa^E_\ell=0$ and
$\mu_{2\ell}^E=\mu_{1\ell}^E$, for all $E$ and $\ell$. Hence ${\cal K}_\ell$
can be identified with the zero operator and $M_{2\ell}=M_{1\ell}=:M_\ell$.
Furthermore, one can check that (\ref{condi}) implies $(\tilde\Q^E)^{3-
\delta(m_+,m_-)}=0$. Therefore, the transformed generators satisfy
	\bea
	&&(\tilde Q_j^2-M_1)(\tilde Q_j^2-M_2)\cdots(\tilde Q_j^2-M_{m_-})
	\tilde Q_j^{1-\delta(m_+,m_-)}=0 \;,
	\label{sa-1}\\
	&&
	\tilde\Q^{3-\delta(m_+,m_-)}=0\;.
	\label{sa-2}
	\eea
We shall term such $\Z_2$-graded TSs, the {\em special} $\Z_2$-graded TSs.

\subsection{$\Z_2$-graded TSs with one Self-Adjoint Generator}

The above analysis of $\Z_2$-graded TSs can be easily applied to the case
where there is a single self-adjoint generator $Q$. In fact, one can read off the
corresponding algebra from Eqs.~(\ref{a1}) and (\ref{a2}). The result is
	\be
	(Q^2-M_1)(Q^2-M_2)\cdots(Q^2-M_{m_-})
	Q^{1-\delta(m_+,m_-)}=0 \;,
	\label{1/2}
	\ee
where $M_\ell$ are self-adjoint operators commuting with $H,~Q$ and $\tau$.
For $m_\pm=1$, this equation reduces to that of the $N=\frac{1}{2}$
supersymmetry, provided that we identify $M_1$ with the Hamiltonian $H$.

\section{Algebraic Structure of $\Z_n$-Graded TSs}

In the preceding section, we have used Definition~3 and the properties 
(\ref{t1}) -- (\ref{t3}) and (\ref{q1}) of the $\Z_2$-grading operator
$\tau$ to obtain the algebra of the $\Z_2$-graded TSs with one generator. 
The main guideline for postulating these properties were Definition~2
and the known grading structure of supersymmetric systems. 

Similarly, in order to obtain the algebraic structure of the $\Z_n$-graded TSs,
with $n>2$, we must first postulate the existence of an appropriate 
$\Z_n$-grading operator. In view of Definitions~1,~2 and~3, such a grading 
operator  $\tau$ must commute with the Hamiltonian --- Eq.~(\ref{t3}) holds ---
and have $n$-distinct eigenvalues $c_\ell$ with eigenspaces ${\cal H}_\ell$ 
satisfying ${\cal H}={\cal H}_1\oplus{\cal H}_2\oplus\cdots{\cal H}_n$. 
The simplest choice for $c_\ell$ are the $n$-th roots of unity, e.g., $c_\ell= q^\ell$
where $q=e^{2\pi i/n}$. This choice suggests the following generalization of
Eqs.~(\ref{t1}) and (\ref{t2}).
	\bea
	\tau^n&=&1\;
	\label{n-t1}\\
	\tau^\dagger&=&\tau^{-1}
	\label{n-t2}
	\eea
Note that for $n=2$, according to Eq.~(\ref{t1}), $\tau^{-1}=\tau$ and
Eq.~(\ref{n-t2}) coincides  with (\ref{t2}). 

In the following we shall only consider  $\Z_n$-graded TSs with one
symmetry generator $\Q$. In order to proceed along the same lines as in
the case of $n=2$, we need to impose a grading condition on $\Q$ similar to
(\ref{q1}). We first consider the simplest case, namely  $\Z_n$-graded UTS
of type $(1,1,\cdots,1)$. 

\subsection{$\Z_n$-Graded UTS of Type $(1,1,\cdots,1)$}
 
For $n=2$, condition (\ref{q1}) implies that the action of $\Q$ on a definite color 
(parity) state vector changes its color (parity) -- a bosonic state changes to a 
fermionic state and vice versa. The simplest generalization of this statement to the 
case $n>2$ is that the action of $\Q$ must change the color of a definite color state 
by one unit, i.e., 
	\be
	\tau\psi=c_\ell\psi~~~{\rm implies}~~~\tau(\Q\psi)=c_{\ell+1}Q\psi\;.
	\label{condi-t}
	\ee
This condition is consistent with Eq.~(\ref{n-t1}). If we impose this condition
on the representations $\Q^E$ and $\tau^E$ in ${\cal H}_E$, then in a basis in 
which $\tau^E$ is diagonal we have
	\be
	\tau^E={\rm diag}(q,q^2,\cdots,q^{n-1},q^n=1)\;,
	\label{n-rep-t}
        \ee
and
        \be
	\Q^E=\left(\begin{array}{ccccccc}
	0&0&0&\cdots&0&0&a_n\\
	a_1&0&0&\cdots&0&0&0\\
	0&a_2&0&\cdots&0&0&0\\
	\vdots&\vdots&\ddots&&\vdots&\vdots&\vdots\\
	0&0&0&\cdots&a_{n-2}&0&0\\
	0&0&0&\cdots&0&a_{n-1}&0\end{array}\right)\;,
	\label{n-rep-q}
	\ee
where $a_\ell$ are complex numbers depending on $E$. 

A simple calculation shows that $\tau^E$ and $\Q^E$ $q$-commute, i.e.
$[\tau^E,\Q^E]_q=0$, where the $q$-commutator is defined by $[O_1,O_2]_q:=
O_1O_2-qO_2O_1$. Generalizing this property of $\tau^E$ and $\Q^E$ to
$\tau$ and $\Q$, we find
	\be
	[\tau,\Q]_q=0\;.
	\label{n-q1}
	\ee
This relation is the algebraic expression of the condition (\ref{condi-t}). It
reduces to Eq.~(\ref{q1}) for $n=2$.

Another consequence of Eq.~(\ref{n-rep-q}) is that a symmetry generator of
a $\Z_n$-graded UTS of type $(1,1,\cdots,1)$ with $n>2$ which satisfies
(\ref{condi-t}) cannot be self-adjoint. Furthermore, one can easily check that 
	\be
	(\Q^E)^n =a_1 a_2 \cdots a_n I_{n \times n }\;.
	\label{4.10}
	\ee
We can generalize this equation to the whole Hilbert space and write it in the
operator form:
           \be
	\Q^n={\cal K}\;.
	\label{4.14}
	\ee
Here ${\cal K}$ is an operator that commutes with all other operators in the algebra.

Next, we seek for the algebraic relations satisfied by the self-adjoint generators:
	\be
	Q_1:={1 \over \sqrt{2}}(\Q+\Q^\dagger)~~~{\rm and}~~~
	Q_2:={-i \over \sqrt{2}}(\Q-\Q^\dagger)\;.
	\label{4.15}
	\ee

Using the results reported in the appendix, namely Eq.~(\ref{xa9}), one can show 
that in an energy eigenspace ${\cal H}_E$, with $E>0$, $Q_1$ and $Q_2$ satisfy
	\bea
	(Q_1^E)^n +\alpha_{n-2} (Q_1^E)^{n-2} 
	+\cdots & =&({1\over \sqrt 2})^n R_1\;,
	\label{4.16}\\
	(Q_2^E)^n +\alpha_{n-2} (Q_2^E)^{n-2} 
	+\cdots & =&({1\over \sqrt 2})^n R_2\;,
	\label{4.17}
	\eea
where $\alpha_\ell$'s are functions of $|a_i|^2$ and $R_1$ and $R_2$ are defined
by
	\[ R_1:=\prod_{k=1}^n a_k +\prod_{k=1}^n a_k^*\;,~~~~
           R_2:=\prod_{k=1}^n(-i a_k) +\prod_{k=1}^n(i a_k^*)\;.\]
Eqs.~(\ref{4.16}) and (\ref{4.17}) can be written in the operator form according to
	\bea
	Q_1^n +M_{n-2} Q_1^{n-2} +\cdots &
	=&({1\over \sqrt 2})^n ({\cal K}+{\cal K}^\dagger)\;,
	\label{4.18}\\ 
	Q_2^n +M_{n-2} Q_2^{n-2} +\cdots &
	=&({1\over \sqrt 2})^n (i^n {\cal K}^\dagger+(-i)^n{\cal K}) \;,
	\label{4.19}
	\eea
where $M_i$s are self-adjoint operators commuting with all other operators.

We can rewrite Eqs.~(\ref{4.18}) and (\ref{4.19}) in the following more symmetric way.
        \begin{itemize}
        \item[~] For $n=2p$,
	\be
        \begin{array}{c}
	(Q_1^2 -{\cal M}_1) (Q_1^2 -{\cal M}_2) \cdots (Q_1^2 -{\cal M}_p)=
	({1 \over 2})^p ({\cal K}+{\cal K}^\dagger)\;,\\
	(Q_2^2 -{\cal M}_1) (Q_2^2 -{\cal M}_2) \cdots (Q_2^2 -{\cal M}_p)=
	({-1 \over 2})^p({\cal K}+{\cal K}^\dagger)\;.\end{array}
	\label{4.20}
	\ee
        \item[~] For $n=2p+1$,
	\be
        \begin{array}{c}
	(Q_1^2 -{\cal M}_1) (Q_1^2 -{\cal M}_2) \cdots (Q_1^2 -{\cal M}_p) Q_1=
	({1 \over \sqrt 2})^{2p+1} ({\cal K}+{\cal K}^\dagger)\;,\\
        (Q_2^2 -{\cal M}_1) (Q_2^2 -{\cal M}_2) \cdots (Q_2^2 -{\cal M}_p)Q_2=
        ({i \over \sqrt 2})^{2p+1} ({\cal K}^\dagger-{\cal K})\;.\end{array}
	\label{4.21}
	\ee
        \end{itemize}
Here ${\cal M}_i$ are also operators that commute with all other operators. Note 
also that for even $p$'s, the algebraic relations for $Q_1$ and $Q_2$ coincide.

Eqs.~(\ref{4.14}), (\ref{4.20}), and (\ref{4.21}) are the defining equations of the 
algebra of $\Z_n$-graded TS of type $(1,1,\cdots,1)$. For $n=2$, they reduce to the familiar
algebra of $\Z_2$-graded TS of type $(1,1)$. In this case, as we discussed in 
section~3.1, we can perform a linear transformation on the self-adjoint generators 
that eliminates the operator ${\cal K}$.
 
Next, consider the case where for all $E>0$, ${\cal K}^E\neq 0$ and introduce the
transformed generator ${\tilde  \Q}$ by 
	\be
	\tilde{\Q}^E=\left(\frac{E}{|\kappa^E|}\right)^{1/n} e^{-i \phi^E/n}\Q^E\;,
	\label{4.12}
	\ee
where
        \be 
        \kappa^E:=a_1a_2\cdots a_n~~~ {\rm and}~~~ 
        e^{i\phi^E}:=\frac{\kappa^E}{|\kappa^E|}\;.
        \ee
In view of Eqs.~(\ref{4.10}) and (\ref{4.12}), it is not difficult to see that
        \be
	({\tilde  \Q}^E)^n=E I_n\;,
	\label{4.13}
	\ee
Writing this equation in the operator form, we are led to
	\be
	H=\tilde{\Q}^n\;.
	\label{4.114}
	\ee
This is precisely the algebra of fractional supersymmetry of order $n$ 
\cite{frac,frac2}.

Next, we examine whether Eq.~(\ref{4.14}) guarantees the desired degeneracy 
structure of the $\Z_n$-graded TS of type $(1,1,\cdots,1)$. In order to address this
question, we suppose that the kernel of ${\cal K}$ is a subset of ${\cal H}_0$. 
Then for all $E>0$, $\kappa^E\neq 0$.

In view of Eq.~(\ref{4.14}), the eigenvalues of $\Q^E$ are of the form
$q^\ell(\kappa^E)^{1/n}$, with $\ell\in\{1,2,\cdots,n\}$. Now, let 
$|q^\ell,\nu_\ell\kt$ denote the corresponding eigenvectors, where $\nu_\ell$'s
are degeneracy labels. We can easily show using Eq.~(\ref{n-q1}) that for
all positive integers $s$, $\tau^s|q^\ell,\nu_\ell\kt$ are eigenvectors of $\Q^E$
with eigenvalue $q^{\ell-s}(\kappa^E)^{1/n}$. This is sufficient to conclude
that all the eigenvalues of $\Q^E$ are either nondegenerate or have the same
multiplicity $N_E$. If for all $E>0$, $N_E=1$, then we will have the desired
degeneracy structure of the $\Z_n$-graded UTS of type $(1,1,\cdots,1)$.
If  there are $E>0$ for which $N_E>1$, then we have a nonuniform $\Z_n$-graded
TS of type $(1,1,\cdots,1)$. 
 
We conclude this section by noting that if we put ${\cal K}=0$ in the algebraic 
relations for $\Z_n$-graded UTS of type $(1,1,\cdots,1)$, we will obtain the 
algebraic relations for $\Z_2$-graded TS. This is not so strange, because putting 
${\cal K}=0$ means that one of the $a_i$'s (say $a_n$) in $\Q^E$ is zero. In this
case, as we explain in the appendix, there is a unitary transformation that transforms
$Q^E_1$ and $Q^E_2$ to off block diagonal matrices. In view of the analysis of 
section~3, this implies that the generators of this kind of $\Z_n$-graded UTSs
satisfy the algebra of  $\Z_2$ graded UT.

 \subsection{$\Z_n$-Graded TS of Arbitrary Type $(m_1,m_2,\cdots,m_n)$}

In order to obtain the algebraic structure of the $\Z_n$-graded TS of arbitrary type
$(m_1,m_2,\cdots,m_n)$, we need an appropriate grading operator. We shall adopt
Eqs.~(\ref{t3}), (\ref{n-t1}), (\ref{n-t2}) and (\ref{n-q1}) as the defining conditions
for our $\Z_n$-grading operator. We shall further assume that 
$m_1\leq m_2\leq\cdots\leq m_n$. This ordering can always be achieved by a 
reassignment of the colors. Again we shall confine our attention to the uniform 
$\Z_n$-graded TS. The algebras of uniform and nonuniform TS of the same type
are identical.

Working in an eigenbasis in which $\tau^E$ is diagonal and using Eq.~(\ref{n-q1}),
we have
	\be
	\tau^E={\rm diag}(\underbrace{q,q,\cdots,q}_{m_1 {\rm times}}
	,\underbrace{q^2,q^2,\cdots,q^2}_{m_2 {\rm times}},\cdots,
	\underbrace{q^n,q^n,\cdots,q^n}_{m_n {\rm times}})\;.
	\label{n-rep-t2}
	\ee
In view of this equation and Eq.~(\ref{n-q1}), we obtain the following matrix	
representation for $\Q^E$.
	\be 
	\Q^E=\left(\begin{array}{ccccccc}
	0&0&0&\cdots&0&0&A_n\\
	A_1&0&0&\cdots&0&0&0\\
	0&A_2&0&\cdots&0&0&0\\
	\vdots&\vdots&\ddots&&\vdots&\vdots&\vdots\\
	0&0&0&\cdots&A_{n-2}&0&0\\
	0&0&0&\cdots&0&A_{n-1}&0\end{array}\right)\;,
	\label{n-rep-q2}
	\ee
where $A_\ell$ (with $\ell\in\{1,2,\cdots,n-1\}$) are complex
$m_{\ell+1}\times m_\ell$ matrices, $A_n$ is a complex $m_{1}\times m_n$ 
matrix, and $0$'s are appropriate zero matrices.

Next, we compute the $n$-th power of $\Q$. The result is
	\be
	(\Q^E)^n=\left(\begin{array}{ccccc}
	A_nA_{n-1}\cdots A_2A_1&0&0&\cdots&0\\
	0&A_1A_nA_{n-1}\cdots A_2&0&\cdots&0\\
	\vdots&\vdots&\ddots&\vdots&\vdots\\
	0&0&0&\cdots&A_{n-1}A_{n-2}\cdots A_1A_n
	\end{array}\right)\;.
	\label{Q^n}
	\ee
In order to find the most general algebraic identity satisfied by $\Q$ we appeal
to the following generalization of Lemma~1. The proof is given in the appendix.
	\begin{itemize}
	\item[~] {\bf Lemma~2:} Let $(m_1,m_2,\cdots,m_n)$ be an $n$-tuple
	of positive integers satisfying $m_1\leq m_2\leq\cdots m_n$, $m:=
	\sum_{\ell=1}^nm_\ell$, $\delta$ is the number of times $m_1$ 
	appears in $(m_1,m_2,\cdots,m_n)$, $Q$ is an $m\times m$ matrix of 
	the form (\ref{n-rep-q2}), and ${\cal P}(x)$ is the characteristic polynomial
	of the $m_1\times m_1$ matrix $A_nA_{n-1}\cdots A_2A_1$. Then
	$Q$ satisfies 
	       \be
	       {\cal P}(Q^n)Q^{n-\delta}=0\,.
                \label{lemma2}
                \ee
	\end{itemize}
Substituting $\Q^E$ for $Q$ in Eq.~(\ref{lemma2}) and writing ${\cal P}(x)$ in
terms of its roots $\kappa_k^E$, we find
	\be
	\left[(\Q^E)^n-\kappa^E_1\right]\left[(\Q^E)^n-\kappa^E_2\right]
	\cdots\left[(\Q^E)^n-\kappa^E_{m_1}\right](\Q^E)^{n-\delta}=0\;.
	\label{n-a-e}
	\ee

Next, we introduce the operators ${\cal K}_k$ for $k\in\{1,2,\cdots,m_1\}$
which commute with the Hamiltonian and have the representation:
	\be
	{\cal K}_k^E=\kappa_k^E~I_m
	\label{n-e-rep}
	\ee
in ${\cal H}_E$. In view of Eqs.~(\ref{n-a-e}) and (\ref{n-e-rep}), we obtain
	\be
	(\Q^n-{\cal K}_1)(\Q^n-{\cal K}_2)\cdots(\Q^n-{\cal K}_{m_1})
	\Q^{n-\delta}=0\;.
	\label{n-a}
	\ee
The operators ${\cal K}_k$ have a similar degeneracy structure as the 
Hamiltonian (at least for the positive energy eigenvalues). Therefore, the
Hamiltonian might be expressed as a function of ${\cal K}_k$.

For a $\Z_n$-graded UTS of type $(1,1,\cdots,1)$, $\delta=n$ and Eq.~(\ref{n-a})
reduces to~(\ref{4.14}). However, one can show that in general Eq.~(\ref{n-a})
does not ensure the desired degeneracy structure of the general $\Z_n$-graded TSs.
This is true even for the case $n=2,~m_+=2,~m_-=1$ considered in section~3. In 
general, the $\Z_n$-graded TSs correspond to a special class of symmetries
satisfying (\ref{n-a}). 

Finally, we wish to note that for a general $\Z_n$-graded TS the algebraic
relations satisfied by the self-adjoint generators are extremely complicated.  We have
not been able to express them in a closed form.

\section{Mathematical Interpretation of $\Delta_{i,j}$}

In order to obtain the mathematical interpretation of the invariants $\Delta_{i,j}$ of
TSs, one must express the Hamiltonian in terms of the symmetry generators. This
can be easily done using the defining algebra for the $\Z_n$-graded TSs of 
type $(1,1,\cdots,1)$. In the following, we discuss the mathematical meaning of the
topological invariants associated with these symmetries.
 
We know that for a $\Z_n$-graded TS of type $(1,1,\cdots,1)$ the operator 
${\cal K}$ has a similar degeneracy structure as the Hamiltonian. Therefore, we may
set $H=f({\cal K})$ where $f$ is a function mapping the eigenvalues $\kappa^E$ to 
$E$. Next, suppose that the kernels of ${\cal K}$ and $H$ also coincide. Then
as far as the general properties of the symmetry is concerned, we can confine our 
attention to the special case where ${\cal K}=H$, i.e., the fractional supersymmetry.
Note that in this case, $\Q^n$ is necessarily self-adjoint. Alternatively, we can use
the rescaled symmetry generator $\tilde{\Q}$ that does satisfy $\tilde{\Q}^n=H$.

In order to obtain the mathematical interpretation of the topological invariants
$\Delta_{i,j}$, we use an $n$-component representation of the Hilbert space in 
which a state vector $\psi$ is represented by a column of $n$ colored vectors 
$\psi_\ell\in{\cal H}_\ell$. In this representation the grading operator is diagonal 
and the generator of the symmetry and the Hamiltonian are respectively expressed by 
$n\times n$ matrices of operators according to
	\bea
	\Q&=&\left(\begin{array}{ccccccc}
	0&0&0&\cdots&0&0&D_n\\
	D_1&0&0&\cdots&0&0&0\\
	0&D_2&0&\cdots&0&0&0\\
	\vdots&\cdots&\ddots&&\vdots&\vdots&\vdots\\
	0&0&0&\cdots&D_{n-2}&0&0\\
	0&0&0&\cdots&0&D_{n-1}&0\end{array}\right)\;,
	\label{q=}\\
	H&=&\Q^n=\left(\begin{array}{cccccc}
	H_1&0&0&\cdots&0&0\\
	0&H_2&0&\cdots&0&0\\
	\vdots&\vdots&\ddots&\cdots&\vdots&\vdots\\
	0&0&0&\cdots&0&H_n
	\end{array}\right)\;.
	 \label{n-rep-q-D}
	\eea
Here $D_n:{\cal H}_{n}\to {\cal H}_1$ and $D_\ell:{\cal H}_{\ell}\to
{\cal H}_{\ell+1}$, with $\ell\in\{1,2,\cdots,n-1\}$, are operators and
	\bea
	H_1&:=&D_nD_{n-1}\cdots D_2D_1\;,\nn\\
	H_2&:=&D_1D_nD_{n-1}\cdots D_2\;,\nn\\
	\cdots& &\cdots\nn\\
	H_{n}&:=&D_{n-1}D_{n-2}\cdots D_1D_n\;.\nn
	\eea
The condition that $H$ is self-adjoint takes the form $H_\ell^\dagger=H_\ell$,
or alternatively
	\be
	D_{\sigma(n)}D_{\sigma(n-1)}\cdots D_{\sigma(2)}D_{\sigma(1)}=
	D_{\sigma(1)}^\dagger D_{\sigma(2)}^\dagger\cdots 
	D_{\sigma(n-1)}^\dagger D_{\sigma(n)}^\dagger\;,
	\label{condi-1}
	\ee
for all cyclic permutations  $\sigma$ of $(n,n-1,n-2,\cdots 1)$. In addition, the
assumption that $H$ has a nonnegative spectrum further restricts $D_\ell$. Note
also that the algebraic relations satisfied by the self-adjoint generators also
put restrictions on the choice of the operators $D_\ell$. This is because the operators
$M_i$ appearing in Eqs.~(\ref{4.18}) and (\ref{4.19}) involve $D_\ell$. The
condition that $M_i$ commute with $\Q$ leads to a set of compatibility 
relations among $D_\ell$.

In view of Eq.~(\ref{n-rep-q-D}) and the fact that $n^{(0)}_\ell$ is the dimension
of the kernel of $H_\ell$, we can easily express the invariant $\Delta_{ij}$ in the
form:
	\be
	\Delta_{i,j}= {\rm dim(ker}~H_j) -{\rm dim(ker}~H_i)\;.
	\label{D=}
	\ee

Suppose that the subspaces ${\cal H}_\ell$ are all identified with a fixed 
Hilbert Space. Consider the special case where 
        \be
        D_3=D_4=\cdots=D_n=1,~~~~D_2=D_1^\dagger,
        \label{xcondi-1}
        \ee
and $D_1$ is a Fredholm operator, then $H_1=D_1^\dagger D_1$, $H_2=
D_1D_1^\dagger$, and there is one independent invariant, namely
	\[\Delta_{1,2}={\rm dim(ker}~D_1^\dagger D_1) -{\rm dim(ker}~
	D_1D_1^\dagger)={\rm dim(ker}~D_1) -{\rm dim(ker}~D_1^\dagger)\;.\]
This is just the analytic index of $D_1$. This example shows that the above 
construction has nontrivial solutions. 

In general, the operator $D_\ell$ need not satisfy (\ref{xcondi-1}). They are
however subject to the above-mentioned compatibility relations. In order to 
demonstrate the nature of these relations, we consider $\Z_3$-graded UTS of
type $(1,1,1)$ with the symmetry generator 
        \be
        \Q=\left(\begin{array}{ccc}
                0 & 0 & D_3 \\
                D_1 & 0 & 0 \\
                0 & D_2 & 0 \\
                \end{array}
                \right).
        \label{xec1}
        \ee          
In this case, the algebraic relations (\ref{4.18}) and (\ref{4.19}) satisfied by the 
self-adjoint generators involve a single commuting operator which we denote by
$M$. Assuming that this operator has the form
        \be
        M={1\over 2}\left( \begin{array}{ccc}
                M_1 & 0 & 0 \\
                0  & M_2 & 0 \\
                0  & 0 & M_3 
                \end{array}
                \right)
                \label{xec2}
                \ee
and enforcing
        \be
        \left[\Q, M\right]=0,~~~ Q_1^3+ MQ_1={1\over \sqrt 2}H\;,
        \label{xec3}
        \ee
we find
        \bea
        M_1D_3&=&D_3M_3 \label{xec4.1}\\
        M_2D_1&=&D_1M_1 \label{xec4.2}\\
        M_3D_2&=&D_2M_2 
        \label{xec4.3}
        \eea
One can manipulate these relations to obtain the following compatibility relations
for $D_\ell$.
        \bea
        D_2^\dagger D_2D_1D_3 & = & D_1 D_3 D_2 D_2^\dagger\;, 
        \label{xec5.1} \\
        D_3^\dagger D_3D_2D_1 & = & D_2 D_1 D_3 D_3^\dagger\;, 
        \label{xec5.2}\\
        D_1^\dagger D_1D_3D_2 & = & D_3 D_2 D_1 D_1^\dagger \;.
        \label{xec5.3}
        \eea
Furthermore, under these conditions on $M_\ell$ and $D_\ell$, one can check that
the relation for $Q_2$, i.e., $Q_2^3+ MQ_2=0$, is identically satisfied.

Next, consider the case where one of the $D_i$'s, say $D_3$, is $1$.  Then we can
use Eqs.~(\ref{xec4.1}) -- (\ref{xec4.3}) to express $M_i$ in terms of $D_1$ and 
$D_2$. This yields
        \be
        M_1=M_3=-(D_1^\dagger D_1 + D_2 D_2^\dagger +1)\,,~~~~
        M_2=-(D_2^\dagger D_2 + D_1 D_1^\dagger +1)\,.
        \label{xec6}
        \ee
Note that we can easily satisfy Eqs.~(\ref{xec5.1}) -- (\ref{xec5.3}), if 
$D_2=D_1^\dagger$, i.e., when (\ref{xcondi-1}) is satisfied. In this case,
        \be
        M_1=M_2=-(2D_1^\dagger D_1+1)=-(2H_1+1)\;,~~~~
        M_3=-(2D_1D_1^\dagger+1)=-(2H_2+1)\;,
        \label{x-special}
        \ee
and
        \be
        M=-(H+\frac{1}{2})\,,
        \label{m=h}
        \ee
where we have used  Eqs.~(\ref{n-rep-q-D}) and (\ref{xec2}).

\section{Examples}
In this section, we examine some examples of quantum systems possessing UTSs.

\subsection{A System with UTS of Type $(2,1)$}
Consider the Hamiltonian 
        \be
        H={1 \over 2}(p^2 +x^2)+{1 \over 2} \left(
                                                         \begin{array}{ccc}
                                                         1 & 0 & 0 \\
                                                         0 & 1& 0 \\
                                                         0 & 0 & -1
                                                         \end{array}
                                                         \right),
        \label{xe81}
        \ee
where $x$ and $p$ are respectively the position and momentum operators. This        
Hamiltonian was originally considered in \cite{para}. It is not difficult to show that
it commutes with
        \be
        \Q=\left(
                \begin{array}{ccc}
                0   &   0   & {1 \over \sqrt 2}(p-ix)\\
                0   &   0   &  0   \\
                0   &  {1 \over \sqrt 2}(p+ix) &  0
                \end{array}
        \right).
        \label{xe8}
        \ee
Furthermore, $\Q$, $H$, and the the self-adjoint generators $Q_j$ satisfy
the algebra of the UTS of type $(2,1)$, i.e.,
        \[Q_1^3=HQ_1\,,~~~~ Q_2^3=HQ_2\,,~~~~ \Q^3= 0\,.\]
      
For this system the grading operator is given by $\tau={\rm diag}(1,1,-1)$; there is
a nondegenerate zero-energy ground state; and the positive energy eigenvalues are 
triply degenerate. Therefore, this system has a UTS of type $(2,1)$.

If we denote by $|n\kt$ the normalized energy eigenvectors of the harmonic oscillator with
unit mass and frequency, then a set of eigenvectors of the Hamiltonian (\ref{xe81})
are given by
        \be
        |\phi_0\kt=\left( \begin{array}{c} 0\\ 0 \\ |0\kt \end{array}\right),
        \label{xe82.9.1}
        \ee
for $E=0$, and 
        \be
          |\phi_n,1\kt= \left( \begin{array}{c} |n-1\kt\\ 0 \\ 0 \end{array}\right) ,~
        |\phi_n,2\kt= \left( \begin{array}{c} 0\\ |n-1\kt \\ 0 \end{array}\right) , ~
        |\phi_n,3\kt= \left( \begin{array}{c} 0\\ 0 \\ |n\kt \end{array}\right),
        \label{xe83.1}
        \ee
for $E=n>0$.

Next, we check the action of the symmetry generator and the grading operator on 
$|\phi_0\kt$ and $|\phi_n,a\kt$. This yields
        \bea
        &&\Q|\phi_0\kt=0,~~~~\tau|\phi_0\kt= - |\phi_0\kt,
        \label{xe8.9}\\
        &&\Q|\phi_n,1\kt=0,~~~~
        \Q|\phi_n,2\kt\rightarrow|\phi_n,3\kt,~~~~
        \Q|\phi_n,3\kt\rightarrow|\phi_n,1\kt\,,
        \label{xe9}\\
        &&\tau|\phi_n,1\kt=|\phi_n,1\kt,~~~~
        \tau|\phi_n,2\kt=|\phi_n,2\kt,~~~~
        \tau|\phi_n,3\kt=-|\phi_n,3\kt.
        \label{xe10}
        \eea
Here we have used $\rightarrow$ to denote equality up to a nonzero multiplicative
constant.

As one can see from Eqs.~(\ref{xe8.9}) and (\ref{xe10}), the ground state has
negative parity and the positive energy levels consist of two positive and one
negative parity states. The topological invariants for this system are $\Delta_{2,1}=
-\Delta_{1,2}=-2$.

\subsection{A System with UTS of Type $(1,1,1)$}
Consider the Hamiltonian 
        \be
        H=\Q^3={1 \over 2}(p^2 +x^2)+{1 \over 2} \left(
                                                         \begin{array}{ccc}
                                                         1 & 0 & 0 \\
                                                         0 & -1& 0 \\
                                                         0 & 0 & 1
                                                         \end{array}
                                                         \right),
        \label{xe2}
        \ee
which is precisely the Hamiltonian~(\ref{xe81}) written in another basis.%
\footnote{We have used some of the results of \cite{frac2} to obtain this 
Hamiltonian.} It possesses a symmetry generated by 
        \be
        \Q=\left(
                \begin{array}{ccc}
                0                        &   0                      &  1\\
                {1 \over \sqrt 2}(p+ix) &   0                      &  0\\
                0                        &  {1 \over \sqrt 2}(p-ix) &  0
                \end{array}
        \right).
        \label{xe1}
        \ee
In fact, it is not difficult to show that
        \be
        \Q^3=H\,.
        \label{xe2.11}
        \ee
Furthermore, one can check that the self-adjoint generators $Q_j$ satisfy
        \be
        Q_1^3+MQ_1=\frac{1}{\sqrt 2}\,H\,,~~~
        Q_2^3+MQ_2=0\,,
        \label{xe2.1}
        \ee
where the operator $M$ is given by
        \be
        M=-\frac{1}{2}\,(p^2+x^2)+\left(\begin{array}{ccc} 
                                                -1&0&0\\
                                                0&0&0\\
                                                0&0&-1\end{array}\right)=
                                                -(H+\frac{1}{2})\;.
        \label{xe2.2}
        \ee
This is in agreement with the more general treatment of section~5. In particular,
Eq.~(\ref{xe2.2}) is a special case of Eq.~(\ref{m=h}).
 
Obviously, $M$ commutes with $H$ and $\Q$. In view of this observation and 
Eqs.~(\ref{xe2.11}), (\ref{xe2.1}), (\ref{4.14}), (\ref{4.18}), and (\ref{4.19}) we can 
conclude that this system has a $\Z_3$-graded UTS of type $(1,1,1)$.

A complete set of eigenvectors of the Hamiltonian (\ref{xe81}) are given by
        \be
        |\phi_0\kt=\left( \begin{array}{c} 0\\ |0\kt \\ 0 \end{array} \right),
        \label{xe82.9}
        \ee
for $E=0$,and 
        \be
        |\phi_n,1\kt= \left( \begin{array}{c} |n-1\kt\\ 0 \\ 0 \end{array}\right) ,~
        |\phi_n,2\kt= \left( \begin{array}{c} 0\\ |n\kt \\ 0 \end{array}\right) , ~
        |\phi_n,3\kt= \left( \begin{array}{c} 0\\ 0 \\ |n-1\kt \end{array}\right),
        \label{xe4}
        \ee
for $E=n>0$.
              
The grading operator is $\tau={\rm diag}(q,q^2,1)$, where $q:=e^{2\pi i/3}$. The 
symmetry generator $\Q$ and the grading operator $\tau$ transform the energy 
eigenvectors according to
        \bea
        \Q|\phi_0\kt=0\,,&&\tau|\phi_0\kt=q^2|\phi_0\kt\,,
        \label{xe4.9}\\
        \Q|\phi_n,1\kt\rightarrow|\phi_n,2\kt,&&
        \Q|\phi_n,2\kt\rightarrow|\phi_n,3\kt,~~~
        \Q|\phi_n,3\kt\rightarrow|\phi_n,1\kt.
        \label{xe5}\\
        \tau|\phi_n,1\kt=q|\phi_n,1\kt,&&
        \tau|\phi_n,2\kt=q^2|\phi_n,2\kt,~~~
        \tau|\phi_n,3\kt=|\phi_n,3\kt.
        \label{xe6}
        \eea
In particular, $|\phi_0\kt$ and $|\phi_n,a\kt$ have colors $q^2$ and $q^a$, 
respectively, and the topological invariants of the system are given by
        \be
        \Delta_{1,2}=-\Delta_{2,1}=1,~~~\Delta_{2,3}=-\Delta_{3,2}=-1,~~~
        \Delta_{1,3}=-\Delta_{3,1}=0.
        \label{xe7}
        \ee

\subsection{A system with UTS of type$(1,1,\cdots,1)$}
Consider the Hamiltonian
	\be
        H={1 \over 2}(p^2 +x^2)+{1 \over 2}\,
        {\rm diag}(\underbrace{1,1,\cdots,1}_{\rm n-1 times},-1)\;.
        \label{xe61}
        \ee
This Hamiltonian has a symmetry generated by 
        \be
        \Q=\left(
                \begin{array}{ccccc}
                0  & 0 & \cdots & 0 &{1 \over \sqrt 2}(p-ix)\\
                1 & 0 & \cdots & 0 & 0 \\
                0 & 1 & \cdots & 0 & 0 \\
                \vdots & \vdots & \ddots & \vdots & \vdots \\
                0 & 0 & \cdots & {1 \over \sqrt 2}(p+ix) &   0
                \end{array} \right)\;,
        \label{xe62}
        \ee
because $\Q^n=H$. 

One can also check that the self-adjoint generators $Q_j$ satisfy 
        \bea
	Q_1^n + M_{n-2} Q_1^{n-2} +\cdots &
	=&({1\over \sqrt 2})^n (2H)\;,\label{xe63}\\
	Q_2^n +M_{n-2} Q_2^{n-2} +\cdots &
	=&({i\over \sqrt 2})^n (1+(-1)^n )H \;,
	\label{xe64}
	\eea
where $M_{n-2k}$ are given by        
        \be 
        M_{n-2k}=(-1)^k\left[{1 \over {2^k}} {{n-k-1} \choose k}+
        {1 \over {2^{k-1}}}{{n-k-1} \choose {k-1}}H\right]\;,
        \label{xe65}
        \ee
and $\left(\begin{array}{c}a\\b\end{array}\right):=\frac{a!}{b!(a-b)!}$.
 
It is not difficult to see that this system has a zero-energy ground state and that the
positive energy levels are $n$-fold degenerate. A complete set of energy 
eigenvectors are given by
        \be
         |0\kt=\left(\begin{array}{c}0 \\ 0 \\ \vdots\\ 0 \\ |0\kt \end{array}\right),
        \label{xe66}
        \ee
for $E=0$, and 
        \be
        |m,1\kt=\left(\begin{array}{c}|m-1\kt \\ 0 \\ \vdots \\ 0 \\ 0 \end{array}\right)
        , \cdots ,
        |m,n-1\kt=\left(\begin{array}{c} 0 \\ 0 \\ \vdots \\ |m-1\kt \\ 0 \end{array}\right)
        ,~~~
        |m,n\kt=\left(\begin{array}{c} 0 \\ 0 \\ \vdots \\ 0 \\  |m\kt \end{array}\right),
        \label{xe67}
        \ee
for $E=m>0$:
       
These observations indicate that the quantum system defined by the Hamiltonian
(\ref{xe61}) has a $\Z_n$-graded UTS of type $(1,1,\cdots,1)$.  Clearly, the
grading operator is $\tau={\rm diag}(q,q^2,\cdots,q^{n-1},q^n=1)$, the ground 
state has color $c_n=q^n=1$, and the nonvanishing topological invariants are 
$\Delta_{\ell,n}=-\Delta_{n,\ell}=1$, where $\ell\in\{1,2,\cdots,n-1\}$.
 
Next, we wish to comment that if we change the sign of the term involving the
matrix ${\rm diag}(\underbrace{1,1,\cdots,1}_{\rm n-1 times},-1)$ in the 
Hamiltonian (\ref{xe61}), then we obtain another quantum system with a 
$\Z_n$-graded UTS of type $(1,1,\cdots,1)$ that is generated by 
        \be
        \Q=\left(
                \begin{array}{ccccc}
                0  & 0 & \cdots & 0 &{1 \over \sqrt 2}(p+ix)\\
                1 & 0 & \cdots & 0 & 0 \\
                0 & 1 & \cdots & 0 & 0 \\
                \vdots & \vdots & \ddots & \vdots & \vdots \\
                0 & 0 & \cdots & {1 \over \sqrt 2}(p-ix) &   0
                \end{array} \right)\;,
        \label{xe62.1}
        \ee
This system has an $(n-1)$-fold degenerate zero-energy ground state.

\section{Conclusions}
We have introduced a generalization of supersymmetry that shares its topological
properties. We gave a complete description of the underlying algebraic structure
and commented on the meaning of the corresponding topological invariants.

We showed that the algebras of the $\Z_2$-graded TSs of type $(m_+,1)$ coincide 
with the algebras of supersymmetry or parasupersymmetry of order $p=2$. The 
algebraic relations obtained for the $\Z_2$-graded TSs of type $(m_+,m_-)$ with 
$m_->1$ include as special cases the algebras of higher order parasupersymmetry 
advocated by Durand and Vinet \cite{durand-vinet}. We also pointed out that the 
algebra of $\Z_n$-graded TS of type $(1,1,\cdots,1)$ is related to the algebra of 
fractional supersymmetry of order $n$. 

Our approach in developing the concept of a TS differs from those of the other
generalizations of supersymmetry in the sense that we introduce TSs in terms of
certain requirements on the spectral degeneracy properties of the corresponding
quantum systems, whereas in the other generalizations of supersymmetry such as
parasupersymmetry and fractional supersymmetry one starts with certain defining
algebraic relations. These relations are usually obtained by generalizing the 
relations satisfied by the generators of symmetries that relate degrees of freedom
with different statistical properties in certain simple models. For example the
Robakov-Spiridonov algebra of parasupersymmetry of order $p=2$ was originally 
obtained by generalizing the algebra of symmetry generators of an oscillator 
involving a bosonic and a ($p=2$) parafermionic degree of freedom \cite{ru-sp}.
In order to investigate the topological content of these (statistical) 
generalizations of supersymmetry, one is forced to study the spectral degeneracy
structure of the corresponding systems. The derivation of the degeneracy structure
using the defining algebraic relations is usually a difficult task. In fact, for
parasupersymmetries of order $p>2$ this problem has not yet been solved. Even for
the parasupersymmetries of order $p=2$ the solution requires a quite lengthy 
analysis \cite{ijmpa-96a}, and the defining algebra does not guarantee the 
existence of any topological invariants. This in turn raises the question of 
the classification of the $p=2$ parasupersymmetries that do have topological 
properties similar to supersymmetry \cite{ijmpa-97}. The analysis of the 
topological aspects of supersymmetry and $p=2$ parasupersymmetry shows that the
information about the topological properties is contained in the spectral 
degeneracy structure of the corresponding systems. This is the main justification
for our definition of a TS. 

Like any other quantum mechanical symmetry, a TS also possesses an underlying 
operator algebra. This algebra contains more practical information about the 
systems possessing the symmetry. As we showed in the preceding sections, the operator
algebras associated with TSs can be obtained using the defining conditions on the
spectral degeneracy structure of these systems. This observation may be viewed as
another indication that, as far as the topological aspects are concerned, the 
spectral degeneracy structure is more basic than the algebraic structure. 

\section*{Acknowledgments}
A.~M.\ wishes to thank Bryce DeWitt for introducing him to supersymmetric quantum
mechanics and encouraging him to work on the topological aspects of supersymmetry
more than ten years ago. We would also like to thank M.~Khorrami and F.~Loran for
interesting discussions and fruitful comments. 

\section*{Appendix}
In this appendix we give the proofs of some of the mathematical results we use in 
sections~3 and 4. In the following we shall denote the characteristic polynomial of a
matrix $M$ by ${\cal P}_M(x)$, i.e., ${\cal P}_M(x)=\det(xI-M)$.

\begin{itemize}
\item[~] {\bf Lemma~0:} Let $X$ and $Y$ be $m\times n$ and $n\times m$ 
matrices respectively. Then 
	\begin{equation}
	{\cal P}_{YX}(\lambda)=\lambda^{n-m}{\cal P}_{XY}(\lambda)\,.
	\label{l1}
	\end{equation}
\item[~] {\bf Proof:} Let
	\begin{equation}
	M:=\left( \begin{array}{cc}
		I_{m} & X \\
		Y & \lambda I_{n}
		\end{array} \right).
	\label{l2}
	\end{equation}
Then using the well-known properties of the determinant, we have	
	\bea
	\det(M)&=&\det \left( \begin{array}{cc}
		I_{m} & X \\
		Y & \lambda I_{n} \end{array} \right)\nn\\
	&=&\det \left( \begin{array}{cc}
		I_{m} & X \\
		0 & \lambda I_{n} -YX
		\end{array} \right)\nn\\
	&=&det( \lambda I_{n} -YX)\nn \\
            &=&{\cal P}_{YX}(\lambda)\;.
	\label{l4}
	\eea
Similarly, we can show that
	\begin{eqnarray}
	\det(M) &=& \det \left( 
		\begin{array}{cc}
		I_{m} & X \\
		Y & \lambda I_{n }
		\end{array} \right)\nn\\
	&=&\lambda^n \det \left( \begin{array}{cc}
		I_{m} & X \\
		{1\over \lambda}Y & I_{n}
		\end{array} \right) \nonumber\\
	&=&\lambda^{n-m} \det \left( \begin{array}{cc}
		\lambda I_{m} &\lambda  X \\
		{1\over \lambda} Y & I_{n}
		\end{array} \right)\nn\\
	&=& \lambda^{n-m}det \left( \begin{array}{cc}
		\lambda I_{m}-XY & 0 \\
		{1\over \lambda}Y & I_{n}
		\end{array} \right) \nonumber\\ 
	&=& \lambda^{n-m}\det(\lambda I_{m} -XY)\nn\\
	&=& \lambda^{n-m}{\cal P}_{XY}(\lambda)\;.
	\label{l5}
	\end{eqnarray}
Eqs.~(\ref{l4}) and (\ref{l5}) yield the identity (\ref{l1}).~$\Box$
\item[~] {\bf Lemma~1:} Let $m_{\pm}$ be positive integers, $m=m_++m_-$, 
and $Q$ be an $m\times m$ matrix of the form
	\begin{equation}
	Q=\left( \begin{array}{cc}
		0 & X \\
		Y & 0
		\end{array} \right),
	\label{xt1}
	\end{equation}
where $X$ and $Y$ are $m_+ \times m_-$ and $m_- \times m_+$ complex 
matrices. Then ${\cal P}_{XY}(Q^2)Q={\cal P}_{YX}(Q^2)Q=0$.
Furthermore, if $m_+=m_-$, then  ${\cal P}_{XY}(Q^2)={\cal P}_{YX}(Q^2)=0$.
\item[~] {\bf Proof:} Let $a_k$ denote the coefficients of ${\cal P}_{YX}(x)$, i.e., 
${\cal P}_{YX}(x)=\sum^n_{k=0} a_k x^k$. According to the Cayley-Hamilton 
theorem,
	\begin{equation}
	{\cal P}_{YX}(YX)=\sum^n_{k=0} a_k(YX)^k =0.
	\label{xt2}
	\end{equation}
In view of this identity, we can easily show that for any positive integer $k$,
	\bea
	Q^{2k}& =& \left( 
		\begin{array}{cc}
		(XY)^k & 0 \\
		0 & (YX)^k
		\end{array}
	\right), \label{xt3}\\
	{\cal P}_{YX}(Q^2)&=&\sum^n_{k=0} a_kQ^{2k}=\left( 
	\begin{array}{cc}
		\sum^n_{k=0} a_k(XY)^k & 0 \\
		0 & \sum^n_{k=0} a_k(YX)^k
		\end{array}
	\right)\nonumber\\
	&=& \left( 
		\begin{array}{cc}
		\sum^n_{k=0} a_k(XY)^k & 0 \\
		0 & 0
	\end{array}
	\right)\label{xt4}\\
	{\cal P}_{YX}(Q^2)Q&=&\left( 
		\begin{array}{cc}
		0 & \sum^n_{k=0} a_k(XY)^k X \\
		0 & 0
		\end{array} \right)\nn\\
	&=&\left( 
		\begin{array}{cc}
		0 & X\sum^n_{k=0} a_k(YX)^k  \\
		0 & 0
		\end{array}
	\right)=0. 
	\eea
A similar calculation yields ${\cal P}_{XY}(Q^2)Q=0$. 
Furthermore, using Lemma~0 one can see that if $m_+=m_-$, 
then ${\cal P}_{XY}(x)={\cal P}_{YX}(x)$. In view of this identity and 
Eq.~(\ref{xt4}), we have (for the case $m_+=m_-$)
${\cal P}_{XY}(Q^2)={\cal P}_{YX}(Q^2) =0$.~$\Box$
\item[~] {\bf Corollary~1:} Consider the $n \times n$ matrix $M$ whose elements 
are given by
	\be	M_{ij}:=a_j \delta_{i , j+1} +a_i^* \delta_{i+1 , j}\;,~~~ 
	i,j\in\{1,2,\cdots , n\}
	\label{xa1}
	\ee
i.e.,
	\be	
	M:=\left( \begin{array}{cccccc}
		0        & a_1^*& 0       & \cdots  & 0      &   0     \\
		a_1     & 0      & a_2^* & \cdots  & 0      &   0     \\
		0        & a_2   & 0       & \cdots & 0      &   0      \\
		\vdots  &\vdots  & \vdots & \ddots & \vdots &\vdots\\
		0        &0       & 0       & \cdots & 0      & a_{n-1}^*\\
		0        &0       & 0       & \cdots & a_{n-1}& 0   
		\end{array}
	\right)
	\label{xa2}
	\ee
Then the characteristic polynomial of $M$ is given by
        \be 
        {\cal P}_M(x)= \cases{{\cal P}_{AA^\dagger}(x^2) & for $n=2p$\cr
	{\cal P}_{AA^\dagger}(x^2)x & for $n=2p+1$,\cr}
	\label{xa3}
	\ee
where $A$ is the matrix with entries
       	\be
	A_{ij}=a_{2i-1}\delta_{ij}+a_{2i}^* \delta_{i+1,j}\;.
	\label{xa4}
        \ee
It is a $p\times p$ matrix for $n=2p$ and a $p \times (p+1)$ matrix for $n=2p+1$.
\item[~] {\bf Proof:} Consider the following unitary transformation
        \be
        M\to{\tilde M}=U^\dagger M U\;,
        \label{xa5}
        \ee
where $U$ is defined by
        \be
        U_{ij}= \cases{\delta_{i,2j} & for $j \le p$ \cr
                \delta_{i,2j-2p-1} & for $j>p.$\cr}  
        \label{xa6}
        \ee
Substituting this equation in (\ref{xa5}), we find
        \be
        {\tilde M}=\left(
                \begin{array}{cc}
                0          &  A\\
                A^\dagger & 0
                \end{array}
        \right)   \;.
        \label{xa7}
        \ee
Now applying Lemma~1, we obtain Eq.~(\ref{xa3}). Furthermore, the coefficeints of
the characteristic polynomial of $M$ are functions of $|a_i|^2$s only.~$\Box$
\item[~] {\bf Corollary~2:} Consider the $n \times n$ complex matrix
        \be
        Q:=\left( 
		\begin{array}{cccccc}
		0       & a_1^*&     0   & \cdots  &     0   & a_n      \\
		a_1    &    0   & a_2^*  & \cdots  &     0   &   0      \\
 		0       & a_2   &   0     &  \cdots &     0   &   0       \\
		\vdots &\vdots  & \vdots & \ddots  & \vdots  &\vdots     \\
		0       &   0    &    0    &  \cdots &  0      & a_{n-1}^*\\
		a_n^*  &   0    &    0    &  \cdots & a_{n-1}& 0   
		\end{array} \right)\;.
	\label{xa8}
        \ee
Then the characteristic polynomial of $Q$ is given by
	\be
	{\cal P}_Q(x) = -(\prod^n_{k=1}a_k +\prod^n_{k=1} a_k^*)+x^n 
	+\beta_{n-2} x^{n-2} +\cdots+\beta_{n-2k} x^{n-2k}+\cdots
	\label{xa9}
	\ee 
where $\beta_{n-2k}$s are functions of $|a_i|^2$s.
\item[~] {\bf Proof:} A straightforward application of the properties of the
determinant, one can show that
        \be
        {\cal P}_Q(x)=-(\prod^n_{k=1}a_k +\prod^n_{k=1} a_k^*)+
        {\cal P}_M(x)-|a_n|^2 {\cal P}_M'(x)\;,
        \label{xa11}
        \ee
where $M'$ is obtained from $M$ by removing the first and last rows and columns.
Now, since ${\cal P}_M(x)$ is an odd (even) polynomial for odd (even)
$n$, ${\cal P}_Q(x)$ will have the form given by Eq.~(\ref{xa9}).~$\Box$
\item[~] {\bf Lemma~2:}  Let $(m_1,m_2,\cdots,m_n)$ be an $n$-tuple
of positive integers satisfying $m_1\leq m_2\leq\cdots m_n$, $m:=
\sum_{\ell=1}^nm_\ell$, $\delta$ is the number of times $m_1$ 
appears in $(m_1,m_2,\cdots,m_n)$, $Q$ is an $m\times m$ matrix of 
the form (\ref{n-rep-q2}), and ${\cal P}(x)$ is the characteristic polynomial
of the $m_1\times m_1$ matrix $A_nA_{n-1}\cdots A_2A_1$. Then
$Q$ satisfies 
	       \be
	       {\cal P}(Q^n)Q^{n-\delta}=0\,.
                \label{xlemma2}
                \ee
 \item[~] {\bf Sketch of Proof:} The proof of this lemma is very similar to
	the proof of Lemma~1. The idea is to multiply the block diagonal matrix
	${\cal P}(Q^n)$ with a power of $Q$ so that one 	obtains a matrix whose 
	entries have a factor ${\cal P}(A_nA_{n-1}\cdots A_1)$. Then 
	one uses the Cayley-Hamilton theorem to conclude that the resulting 
	matrix must vanish. One can show by inspection that the smallest 
	nonnegative integer $r$ for which a factor of ${\cal P}(A_nA_{n-1}
	\cdots A_1)$ occurs in all the entries of ${\cal P}(Q^n)Q^r$ is 
	$n-\delta$.
	\end{itemize}


\begin{thebibliography}{99}
\bibitem{witten-82} E.~Witten, Nucl.\ Phys.\ B {\bf 202}, 253 (1982). 
\bibitem{susy} L.~E.~Gendenshtein and I.~V.~Krive, Sov.\ Phys.\ Usp.\ 
{\bf 28}, 645-666 (1985)
\bibitem{review} F.~Cooper, A.~Khare, and U.~Sukhatme, Phys.\ Rep.\ 
{\bf 251}, 267-385 (1995).
\bibitem{junker} G.~Junker, {\em Supersymmetric Methods in Quantum and 
Statistical Physics} (Springer-Verlag, Berlin, 1996).
\bibitem{morse} E.~Witten, J.~Diff.~Geom.\ {\bf 17}, 661 (1982).
\bibitem{susy-index} L.~Alvarez-Gaume, Commun.\ Math.\ Phys.\ 
{\bf 90}, 161 (1983);\\
L.~Alvarez-Gaume, J.\ Phys.\ A: Math.\ Gen.\ {\bf 16}, 4177 (1983);\\
P.~Windey, Acta.\ Phys.\ Pol.\ B {\bf 15}, 453 (1984);\\
A.~Mostafazadeh, J.\ Math.\ Phys.\ {\bf 35}, 1095 (1994).
\bibitem{ru-sp} V.~A.~Rubakov and V.~P.~Spiridonov, Mod.\ Phys.\ Lett.\ 
{\bf A 3}, 1337 (1988).
\bibitem{para} J.~Beckers and N.~Debergh, Nucl.\ Phys.\ B {\bf 340}, 
767 (1990).
\bibitem{def} D.~Bonatsos and C.~Daskaloyannis, Phys.\ Lett.\ B 
{\bf 307}, 100 (1993);\\
N.~Debergh, J.\ Phys.\ A: Math.\ Gen.\ {\bf 26}, 7219 (1993);\\
K.~N.~Ilinski and V.~M.~Uzdin, Mod.\ Phys.\ Lett.\ A {\bf 8}, 2657 (1993).
\bibitem{frac} 
C.~Ahn, D.~Bernard, and A.~Leclair, Nucl.\ Phys.~B {\bf 346}, 409 (1990);\\ 
L.~Baulieu and E.~G.~Floratos, Phys.\ Lett.~B {\bf 258}, 171 (1991);\\
R.~Kerner, J.~Math.\ Phys.\ {\bf 33}, 403 (1992);\\
S.~Durand, Mod.\ Phys.\ Lett.\ A {\bf 8}, 1795 (1993);\\
S.~Durand, Mod.\ Phys.\ Lett.\ A {\bf 8}, 2323 (1993);\\
A.~T.~Filippov, A.~P.~Isaev, and R.~D.~Kurdikov, Mod.\ Phys.\ Lett.~A {\bf 7},
2129 (1993);\\
N.~Mohammedi,  Mod.\ Phys.\ Lett.~A {\bf 10}, 1287 (1995);\\
N.~Fleury and M.~Rausch~de~Traubenberg, Mod.\ Phys.\ Lett.~A {\bf 11},
2899 (1996);\\
J.~A.~de~Azc\'arraga and A.~J.~Macfarlane, J.~Math.\ Phys.\ {\bf 37}, 1115 (1996);\\
R.~S.~Dunne, A.~J.~Macfarlane, J.~A.~de~Azc\'arraga, and J.~C.~P\'erez Bueno,
Int.\ J.~Mod.\ Phys.\ A {\bf 12}, 3275 (1997).  
\bibitem{frac2} S.~Durand, Phys.\ Lett.\ B {\bf 312}, 115 (1993).
\bibitem{ex-ge} N.~V.~Borisov,  K.~N.~Ilinski, and V.~M.~Uzdin, Phys.\ Lett.\ A
{\bf 169}, 422 (1992);\\
A.~D.~Dolgallo and K.~N.~Ilinski, Ann.\ Phys.\ {\bf 236}, 219 (1994). 
\bibitem{ijmpa-97} A.~Mostafazadeh, Int.\ J.\ Mod.\ Phys.\ A {\bf 12}, 2725 
(1997).
\bibitem{p34a} A.~Mostafazadeh and K.~Aghababaei Samani, Mod.\ Phys.\ Lett.\
A {\bf 15}, 175 (2000).
\bibitem{ijmpa-96a}  A.~Mostafazadeh, Int.\ J.\ Mod.\ Phys.\ A {\bf 11}, 1057 
(1996).
\bibitem{durand-vinet} S.~Durand and L.~Vinet, Mod.\ Phys.\ Lett.\ A {\bf 6},
3165 (1991).
\end{thebibliography}
\end{document}